\documentclass[epjCONF,columns]{svjour} 
\usepackage{graphics}
\usepackage[varg]{txfonts} 
\usepackage[latin1]{inputenc}

\session-title{NSRT12}
\newcommand{\beq}{\begin{equation}}
\newcommand{\eeq}{\end{equation}}
\newcommand{\bea}{\begin{eqnarray}}
\newcommand{\eea}{\end{eqnarray}}
\newcommand{\eps}{\varepsilon}
\newcommand{\Vef}{{\cal V}_{\rm eff}}
\begin{document}
\title{An  {\it ab initio} theory of double odd-even mass differences in nuclei}

\author {E.\,E. Saperstein\inst{1}\fnmsep\thanks{\email{saper@mbslab.kiae.ru}}
\and M. Baldo\inst{2} \and
N.\,V. Gnezdilov\inst{1} \and U. Lombardo\inst{3} \and S.\,S. Pankratov\inst{1}, \inst{4}  }
\institute{Kurchatov Institute, 123182, Moscow, Russia
\and INFN, Sezione di Catania, 64 Via S.-Sofia, I-95125 Catania, Italy
\and INFN-LNS and University of Catania, 44 Via S.-Sofia, I-95125 Catania, Italy
\and Moscow Institute of Physics and Technology, 123098 Moscow, Russia}
\abstract{
Two aspects of the problem of evaluating double odd-even mass differences $D_2$ in semi-magic nuclei are
studied related to existence of two components with different properties, a superfluid nuclear subsystem
and a non-superfluid one. For the superfluid subsystem, the difference $D_2$ is approximately equal to $2\Delta$,
the gap $\Delta$ being the solution of the gap equation. For the non-superfluid
subsystem,  $D_2$ is found by solving the equation for two-particle Green function
for normal systems. Both equations under consideration contain the same effective
pairing interaction. For the latter,
the semi-microscopic model  is used in which the main term calculated from the first principles is
supplemented with a small phenomenological addendum containing one
phenomenological parameter supposed to be universal for all medium and heavy atomic
nuclei.
}
%
\maketitle
\section{Introduction}
\label{intro}
Recently, a progress has been made in the {\it 'ab initio'}  theory of nuclear pairing,
first, by the Milan group \cite{milan1,milan2,milan3},  a bit
later by Duguet et al. \cite{Dug1,Dug2}, and finally by the
Moscow-Catania group \cite{Pankr1,Bald1,Pankr2,Pankr3,Sap1}.
By {\it 'ab initio'} we
mean a theory of Brueckner type starting from a free $NN$-potential, not a real
{\it ab initio} approach based on the QCD theory. The quotes indicate the fact that,
in any of the cited references, the authors deal with
a calculation which is not completely {\it ab initio} even at such lowest level.
Indeed, the self-consistent mean field  used in each of them is calculated by means
of an Energy Density Functional (EDF) containing phenomenological
parameters. Solving the BCS equation for the gap
 $\Delta$ with a realistic $NN$-potential for the pairing interaction is the main ingredient of
all of them. This is exactly the prescription of the simplest version of the
Brueckner theory as  the ladder diagram summation is already performed
in the gap equation itself.

In the first  paper of the Milan series, the BCS gap equation for
neutrons with the Argonne v$_{14}$ potential was solved for the
nucleus $^{120}$Sn, the Saxon-Woods Shell-Model basis with the bare
neutron mass $m^*=m$. Rather
optimistic result  was obtained for the gap value, $\Delta_{\rm
BCS}=2.2$ MeV which  is bigger, but not dramatically, than the experimental one,
$\Delta_{\rm exp}\simeq 1.3$ MeV, leaving some
 hope of achieving a good agreement by finding
corrections to the scheme. In Refs. \cite{milan2,milan3},
the effective mass $m^*\neq m$ was introduced into the gap equation
within the Skyrme--Hartree--Fock (SHF) method with the Sly4 force \cite{SLy4}. In this case  the effective
mass $m^*(r)$ is space dependent and essentially different from
the bare one $m$. E.g., in nuclear matter the Sly4 effective mass is
equal to $m^*=0.7 m$. In the  weak coupling
limit of the BCS theory, the gap is exponentially dependent, i.e.
$\Delta \propto \exp(1/g)$, on the inverse dimensionless pairing
strength $g= m^*{\cal V}_{\rm eff} k_{\rm F}/\pi^2$, where
${\cal V}_{\rm eff}$ is the effective
pairing interaction. Therefore, a strong suppression of the gap
takes place in the case of $m^*< m$. The value of $\Delta_{\rm
BCS}=0.7$ MeV was obtained in Ref. \cite{milan2} and $\Delta_{\rm
BCS}=1.04$ MeV, in Ref. \cite{milan3}. In both cases, the too small
value of the gap was explained by invoking various many-body
corrections to the BCS approximation. The main correction is due to
the exchange of low-lying surface vibrations (``phonons''),
contributing to the gap around 0.7 MeV \cite{milan2}, so that the
sum  turns out to be $\Delta=1.4$ MeV very close to the experimental
value. In Ref. \cite{milan3}, the contribution of the induced
interaction caused by exchange of the high-lying in-volume
excitations was added either, the sum again is equal to
$\Delta\simeq 1.4$ MeV. Thus, the calculations of Refs.
\cite{milan2,milan3} showed that the effects of $m^*\neq m$ and of
many-body corrections to the BCS theory are necessary  to explain
the difference of ($\Delta_{\rm BCS}-\Delta_{\rm exp}$). In
addition, they have different sign and partially compensate each
other. Unfortunately, both effects contain large uncertainties. This
point was discussed in Refs. \cite{Bald1,Pankr3,Sap1}.

To avoid such uncertainties, a semi-microscopic model for nuclear
pairing was suggested in Refs.
\cite{Pankr2,Pankr3,Sap1}. It starts from the {\it ab initio} BCS
gap equation with the Argonne force v$_{18}$ treated with the
two-step method. The complete Hilbert space $S$ of the problem was
split into the model subspace $S_0$  and the
complementary one $S'$. The gap equation is solved in the model
space with the effective interaction ${\cal V}_{\rm eff}$ which is found
in the complementary subspace in terms of the initial $NN$-potential
${\cal V}$ with $m^*=m$.   This {\it
ab-initio} term of ${\cal V}_{\rm eff}$ was supplemented by a small addendum
proportional to the phenomenological parameter $\gamma$ that should
hopefully embody all corrections to this simplest BCS scheme.  This parameter
is supposed to be the same for all heavy and medium nuclei, either
for neutrons and protons.

The ``experimental'' gap value $\Delta_{\rm exp}$ for semi-magic nuclei is
usually identified with
one half of one of that obtained from the following double odd-even mass differences:
\beq D_{2n}^+(N,Z) =  M(N+2,Z)+ M(N,Z)-2M(N+1,Z),\label{d2npl} \eeq
\beq D_{2n}^-(N,Z) =  -M(N-2,Z)- M(N,Z)+2M(N-1,Z),\label{d2nmi}\eeq
 \beq D_{2p}^+(N,Z) =  M(N,Z+2)+ M(N,Z)-2M(N,Z+1),\label{d2ppl} \eeq
\beq D_{2p}^-(N,Z) = -M(N,Z-2)-M(N,Z)+ 2M(N,Z-1),\label{d2pmi}\eeq
where $Z$ value is magic in Eqs. (\ref{d2npl}),(\ref{d2nmi}) and $N$ is magic
in Eqs. (\ref{d2ppl}),(\ref{d2pmi}).
More explicitly, in Refs. \cite{Pankr3,Sap1} the experimental
gap values were defined as follows:
\beq 2\Delta_n^{\rm exp}(N,Z) = 1/2(D_{2n}^+(N,Z)+D_{2n}^-(N,Z)),\label{deln-exp} \eeq
\beq 2\Delta_p^{\rm exp}(N,Z) = 1/2(D_{2p}^+(N,Z)+D_{2p}^-(N,Z)).\label{delp-exp} \eeq
The accuracy of such a prescription  was estimated in \cite{Pankr3} as
${\simeq}0.1\div0.2\;$MeV. Approximately the same accuracy holds for the ``developed pairing'' approximation
in the gap equation, with conservation of the particle number only on average \cite{part-numb},
used in [1--10].

There is one more physical quantity in semi-magic nuclei which can be evaluated in terms of
the same effective interaction as the pairing gap. This is the set of the same double odd-even
mass differences (\ref{d2npl})--(\ref{d2pmi}), but for the non-superfluid subsystem, so that now
$N$ is magic and $Z$ arbitrary in Eqs. (\ref{d2npl}),(\ref{d2nmi}) and {\it vice versa} in Eqs. (\ref{d2ppl}),(\ref{d2pmi}).
In non-superfluid nuclei, the mass differences, Eqs.
(\ref{d2npl}),(\ref{d2nmi}), coincide with poles in the total energy $E$ plane of the two-particle
Green function $K(1,2,3,4)$ for  normal systems  \cite{AB} in the $nn$-channel,
and Eqs. (\ref{d2ppl}),(\ref{d2pmi}), in the $pp$-channel. The equation for $K$ in the channel $S=0,L=0$
could be expressed in terms of
the same effective interaction $\Vef$  as the pairing gap.
This point was marked  in the old paper \cite{Sap-Tr}, where these differences for double-magic nuclei were analyzed
within the theory of finite Fermi systems (TFFS) \cite{AB}. In this article,
the density dependent effective pairing interaction was introduced for the first time and arguments were found in favor of
of the surface dominance in this interaction.

It is worth to stress that this calculation of mass difference for the non-superfluid subsystem is a more
rigorous operation than the identification with the double gap $\Delta$ in the superfluid one.
The first results of such calculations with the use of the semi-microscopic model for the effective
pairing interaction are presented in \cite{Gnezd1}.

\section{The semi-microscopic model for nuclear pairing}
\label{sec:1}
The general many-body form of the equation for the pairing gap is \cite{AB}
\beq \Delta_{\tau} =  {\cal U}^{\tau} G_{\tau} G^s_{\tau}
\Delta_{\tau}, \label{del} \eeq where $\tau=(n,p)$ is the isotopic
index, ${\cal U}^{\tau}$ is the $NN$-interaction block irreducible
in the two-particle $\tau$-channel, and
 $G_{\tau}$  ($G^s_{\tau}$) is the one-particle Green function without (with)
 pairing. A symbolic multiplication denotes the integration over
energy and intermediate coordinates and summation over spin
variables as well. The BCS approximation in Eq. (\ref{del}) means,
first, the change of the block ${\cal U}$ of irreducible interaction
diagrams with the free $NN$-potential ${\cal V}$
 and, second, the use of simple quasi-particle Green
functions $G$ and $G^s$, i.e. those without phonon corrections and
so on. In this case, Eq. (\ref{del}) iss greatly simplified and
can be reduced to the form typical of the Bogolyubov method, \beq
\Delta_{\tau} = - {\cal V}^{\tau} \varkappa_{\tau}\,, \label{delkap}
\eeq where \beq\varkappa_{\tau}=\int \frac {d\eps}{2\pi i}G_{\tau}
G_{\tau}^s\Delta_{\tau}
 \label{defkap}\eeq is the anomalous density matrix
which can be expressed explicitly in terms of the Bogolyubov
functions $u$ and $v$,
\beq \varkappa_{\tau}({\bf r}_1,{\bf r}_2) = \sum_i u_i^{\tau}({\bf
r}_1) v_i^{\tau}({\bf r}_2). \label{kapuv} \eeq Summation in Eq.
(\ref{kapuv}) scans the complete set of Bogolyubov functions with
eigen-energies $E_i>0$.

  Then we split the complete Hilbert space of the
pairing problem, $S=S_0+S'$, where the model subspace $S_0$ includes
the single-particle states with energies less than a separation
energy $E_0$. The gap equation is
solved in the model space: \beq \Delta_{\tau} = {\cal V}^{\rm BCS}_{\tau,{\rm eff}} G_{\tau}
G^s_{\tau} \Delta_{\tau}|_{S_0}, \label{del0} \eeq with the
effective pairing interaction ${\cal V}^{\rm BCS}_{\tau,{\rm eff}}$ instead of the block
${\cal V}^{\tau}$ in the BCS version of the original gap equation
(\ref{del}). It obeys the Bethe--Goldstone type equation in the
subsidiary space, \beq {\cal V}^{\rm BCS}_{\tau,{\rm eff}} = {\cal V}^{\tau} + {\cal
V}^{\tau} G_{\tau} G_{\tau} {\cal V}^{\rm BCS}_{\tau,{\rm eff}}. \label{Vef} \eeq In
this equation, the pairing effects can be neglected provided the
model space is sufficiently large, $E_0\gg \Delta$. That is why we
replaced the Green function $G^s_{\tau}$ for the superfluid system
with its counterpart $G_{\tau}$ for the normal system.  To solve
 Eq. (\ref{Vef}) in non-homogeneous systems
 a new form of the local approximation, the Local
Potential Approximation (LPA), was developed by the Moscow--Catania
group. Originally, it was developed for semi-infinite nuclear matter, then applied to the slab
of nuclear matter (see review article \cite{Rep}) and, finally, to finite nuclei
\cite{Pankr1,Bald1}. It turned out that, with a very high accuracy,
at each value of the average c.m. coordinate ${\bf R}=({\bf r}_1 +
{\bf r}_2 +{\bf r}_3 +{\bf r}_4)/4$, one can use in Eq. (\ref{Vef})
the formulae  of the infinite system embedded into the constant
potential well $U=U({\bf R})$. This significantly simplifies the
equation for ${\cal V}_{\rm eff}$, in comparison with the original equation for
$\Delta$. As a result, the subspace $S'$ can be chosen as large as
necessary to achieve the convergence. The accuracy of  LPA depends on
the separation energy $E_0$. For finite nuclei, the value of
$E_0{=}40\;$MeV guarantees an accuracy higher than 0.01 MeV for the
gap $\Delta$.

To avoid uncertainties that affect the corrections to
the BCS scheme discussed above, a semi-microscopic model was
suggested in Refs. \cite{Pankr2,Pankr3,Sap1}. In this model, a small
phenomenological addendum to  the effective pairing interaction is
introduced which embodies approximately all these corrections. The
simplest ansatz for it is

\bea {\cal V}^{\tau}_{\rm
eff}({\bf r}_1,{\bf r}_2,{\bf r}_3,{\bf r}_4) & = &
{\cal V}^{\rm BCS}_{\tau,{\rm eff}}({\bf r}_1,{\bf r}_2,{\bf r}_3,{\bf r}_4) \nonumber \\ & + &
\gamma^{\tau} C_0 \frac {\rho(r_1)}{\bar{\rho}(0)}
\prod_{i=2}^4\delta ({\bf r}_1 - {\bf r}_i). \label{Vef1} \eea

Here
$\rho(r)$ is the density of nucleons of the kind under
consideration, and $\gamma^{\tau}$ are dimensionless
phenomenological parameters. To avoid any influence of the shell
fluctuations in the value of ${\rho}(0)$, the average central
density ${\bar{\rho}(0)}$ is used in the denominator of the
additional term. It is averaged over the interval of $r{<}2\;$fm.
The first, {\it ab initio}, term in the r.h.s. of Eq. (\ref{Vef1})
is the solution of  Eq. (\ref{Vef}) in the framework of the LPA
method described above, with $m^*{=}m$ in the subspace $S'$.

In Ref.\cite{Pankr3}, the above equations were solved in the
self-consistent $\lambda$-basis of the
EDF by Fayans et al. \cite{Fay1,Fay}. Two sets of the functional
were used, the original one DF3 \cite{Fay} and its modification
DF3-a \cite{Tol-Sap}. In the latter, the spin-orbit and effective
tensor terms of the original functional were modified. The results
for the pairing gap in three chains of semi-magic nuclei are
displayed in Figs. 1--3.

\begin{figure}
\resizebox{1.00\columnwidth}{!}
{\includegraphics{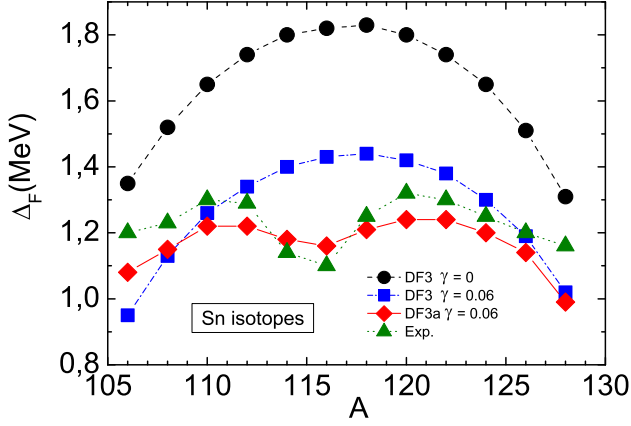}}
\caption{Neutron gap in Sn isotopes }
\label{fig:deln-sn}
\end{figure}

\begin{figure}
\resizebox{1.00\columnwidth}{!}
{\includegraphics {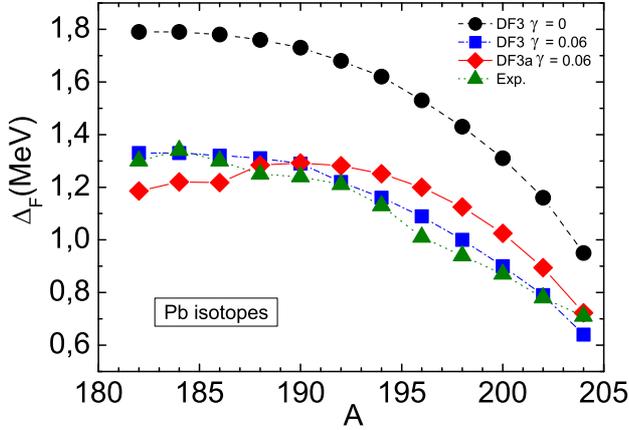}}
\caption{Neutron gap in Pb isotopes  }
\label{fig:deln-pb}
\end{figure}

In accordance with the recipe of Ref. \cite{milan3}, we represent
the theoretical gap with the ``Fermi average'' combination \beq
\Delta_{\rm F}=\sum_{\lambda}{(2j{+}1)\Delta_{\lambda
\lambda}}/\sum_{\lambda}(2j{+}1), \label{DelF}\eeq where the
summation is carried out over the states $\lambda$ in the interval
of $|\eps_{\lambda}{-}\mu|{<}3\;$MeV. The ``experimental'' gap is
determined by the symmetric 5-term odd-even mass differences (\ref{deln-exp}),(\ref{delp-exp}).

Let us begin from neutron pairing and consider first the tin
isotopes, Fig. \ref{fig:deln-sn}. We see that the BCS gap ($\gamma{=}0$) is
approximately 30\% greater than the experimental one. Switching on
the phenomenological addendum with $\gamma{=}0.06$ makes the
theoretical gap values closer to experiment. However, the predictions of the
two versions of the functional used are significantly different,
being much better for the DF3-a functional. In particular, the
$A$-dependence of the experimental gap is reproduced  with a
pronounced minimum in the center of the chain. As the analysis in
Ref.\cite{Pankr3} has shown, this strong difference between results
for two functionals is due to the strong influence on the gap of
the high $j$ intruder state $1h_{11/2}$. Its position depends
essentially on the spin-orbit parameters and is noticeably different
for DF3 and DF3-a functionals.

In the lead chain, see Fig. \ref{fig:deln-pb}, the overall pattern is quite similar.
Again the BCS gap is approximately 30\% bigger of the experimental
one and again the inclusion of the phenomenological term with
$\gamma{=}0.06$ gives a qualitative agreement. Now, the difference
between the two functionals is much smaller. In this case, the agreement is
rather perfect for the DF3 functional and a little worse for the
DF3-a one, but also within limits for the accuracy discussed above.

\begin{figure}
\resizebox{1.00\columnwidth}{!}
 {\includegraphics {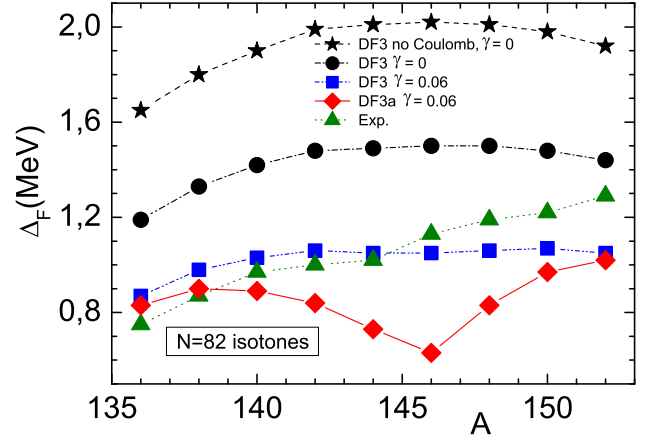}}
\caption{Proton gap for $N=82$ isotones}
\label{fig:del-p}
\end{figure}

Let us go to proton pairing, $N{=}82$ chain, see Fig. \ref{fig:del-p}.  In this
case, the Coulomb interaction should be included into the pairing
effective interaction, \beq  {\cal V}_{\rm eff}^p={\cal V}_{\rm eff}^n+{\cal V}_{\rm C}.\eeq As
it is argued in Ref.\cite{Pankr3}, the bare Coulomb potential could
be used in this equation with high accuracy. The strong Coulomb
effect in the gap is demonstrated in Fig. \ref{fig:del-p}.  It is also explained
with the exponential dependence of the gap on the pairing
interaction. It should be mentioned that Duguet and co-authors
\cite{Dug2} were the first who inserted explicitly the Coulomb interaction
into the pairing force for protons. Only after inclusion of the
Coulomb interaction into ${\cal V}_{\rm eff}$, we can use the same value of
$\gamma=0.06$ for protons and neutrons.  As for tin isotopes, the
difference between DF3 and DF3-a results is rather strong, now in
favor of the DF3 functional. This effect is again due to different
positions of the $1h_{11/2}$ level, but now for protons. Overall
agreement with experiment is for protons worse, maybe, because of
closeness of some nuclei to the region of the phase transition to
deformed state.

 As it is argued in Introduction, we may expect to reach accuracy for the gap of the order of  $\simeq(0.1
 \div 0.2)\;$MeV. In practice, the overall disagreement of the theory with the data is $\sqrt{\overline{(\Delta_{\rm th}-
\Delta_{\rm exp})^2}}{\simeq}0.13\;$MeV for the DF3 functional and
${\simeq}0.14\;$MeV, for the DF3-a one. For the mass differences $D_2$ these numbers
should be multiplied by a factor of two.

\section{Double mass differences in non-superfluid subsystems}
\label{sec:2}
The Lehmann expansion  for the two-particle Green function $K$ in a non-superfluid system reads,
in the single-particle wave functions $|1\rangle{=}|n_1,l_1,j_1,m_1\rangle$ representation, \cite{AB}:
\beq K_{12}^{34}(E)=\sum_s
\frac {\chi^s_{12}\chi^{s*}_{34}} {E-E_s^{+,-} \pm i\gamma},
\label{Lem}\eeq where $E$ is the total energy in the two-particle channel and
$E_s^{+,-}$ denote the eigen-energies of  nuclei with
 two particles and two holes, respectively, added to the original nucleus. They are often interpreted as the ``pair
vibrations'' \cite{BM2}.  Instead of the Green function $K$, it is convenient
to use the two-particle interaction amplitude
$\Gamma$: \beq K = K_0 + K_0 \Gamma K_0, \label{gam}\eeq where
$K_0=GG$. The amplitude
$\Gamma$ obeys the following equation \cite{AB}: \beq \Gamma = {\cal U}+{\cal U} GG
\Gamma, \label{eqgam}\eeq where ${\cal U}$ is the same irreducible interaction block as in Eq. (\ref{del}).
Again, within the Brueckner theory,  the block  ${\cal
U}$ should be replaced with the realistic potential ${\cal V}$
which does not depend on the energy. Then the integration over the relative energy can be readily carried
out in Eq. (\ref{eqgam}): \beq
 A_{12} {=}  \int \frac {d\eps}{2\pi
i}G_1\left(\frac E 2 {+}\eps \right) G_2\left(\frac E 2 {-}\eps
\right)
 {=}\frac {1{-}n_1{-}n_2}
{E{-}\eps_1{-}\eps_2}, \label{Alam} \eeq where $\eps_{1,2}$
are the single-particle energies and $n_{1,2}{=}(0;1)$, the corresponding occupation numbers.
As the result, we obtain: \beq \Gamma = {\cal V}+{\cal V} A
\Gamma. \label{eqgam1}\eeq

Simple transformations in Eq. (\ref{gam}) or (\ref{eqgam1}), in vicinity of the pole  $E{=}E_s^{+,-}$,
lead to the following equation for the eigenfunctions  $\chi^s$: \beq (E_s-\eps_1-\eps_2)
\chi^s_{12}=(1-n_1-n_2) \sum_{34}{\cal V}_{12}^{34}
\chi^s_{34}\label{eqchi}.\eeq It is different from the Shr\"{o}dinger equation for
two interacting particles in an external field only for the factor
$(1-n_1-n_2)$ which reflects the many-body character of the problem, in particular, the Pauli principle.
The direct solution of this equation is  complicated by the same reasons as of the {\it ab initio} BCS
gap equation described in  Sect. 2. The same two-step method is used. The usual renormalization
of Eq. (\ref{eqchi}) transforms it into the analogous equation in the model space:\beq (E_s{-}\eps_1{-}\eps_2)
\chi^s_{12}{=}(1{-}n_1{-}n_2) {\sum_{34}}^0 \left({\cal V}_{\rm eff}\right)_{12}^{34}\; \chi^s_{34},\label{eqchi0}\eeq where
the effective interaction ${\cal V}_{\rm eff}$ coincides with that
of pairing problem, Eq. (\ref{Vef}), provided the same value of the separation energy $E_0$ is used.
The next step consists in the use of  ansatz (\ref{Vef1})
to take into account corrections to the Brueckner theory with a phenomenological addendum
($ \sim \gamma$).

The double mass differences (\ref{d2npl}) -- (\ref{d2pmi}) are identified with the two first solutions
 $E_s^{+,-}$ of Eq. (\ref{eqchi0}), corresponding to the addition of two particles (holes) to the
 magic core into the state $\eps_1=\eps_2=\mu^{+,-}$, where the chemical potentials
$\mu^{+,-}$ are defined in a usual way as mass differences, e.g., $\mu_p^+=E_{\rm B}(N,Z+1)-E_{\rm B}(N,Z)$. Then,
the energy difference in the left-hand side of  Eq. (\ref{eqchi0}) is
$E_s^{+,-}-2\mu^{+,-}=D^{+,-}$. If we are interested only in these solutions, we may rewrite Eq.
(\ref{eqchi0})  as follows: \beq E_s-2\mu{=}(1{-}2n_1)\left(
\Gamma'(E_s)\right)_{11}^{11}, \label{Es}\eeq where
\bea \left(
\Gamma'(E_s)\right)_{12}^{34} & = & \left({\cal V}_{\rm
eff}\right)_{12}^{34} +{\sum_{56}}'\left({\cal V}_{\rm
eff}\right)_{12}^{56} \nonumber \\ & \times & \frac {1{-}n_5{-}n_6}
{E_s {-}\eps_5{-}\eps_6} \left( \Gamma'(E_s)\right)_{56}^{34}.
\label{Gamma1} \eea
The accent on the sum denotes that the two-particle state $5{=}6{=}1$
is excluded. In all  formulas  (\ref{eqchi0})
-- (\ref{Gamma1})  the angular momenta of two-particle states  $|12\rangle$, $|34\rangle$ are coupled to
the total angular momentum  $I{=}0$.

\begin{figure}
\resizebox{1.0\columnwidth}{!}
{\includegraphics{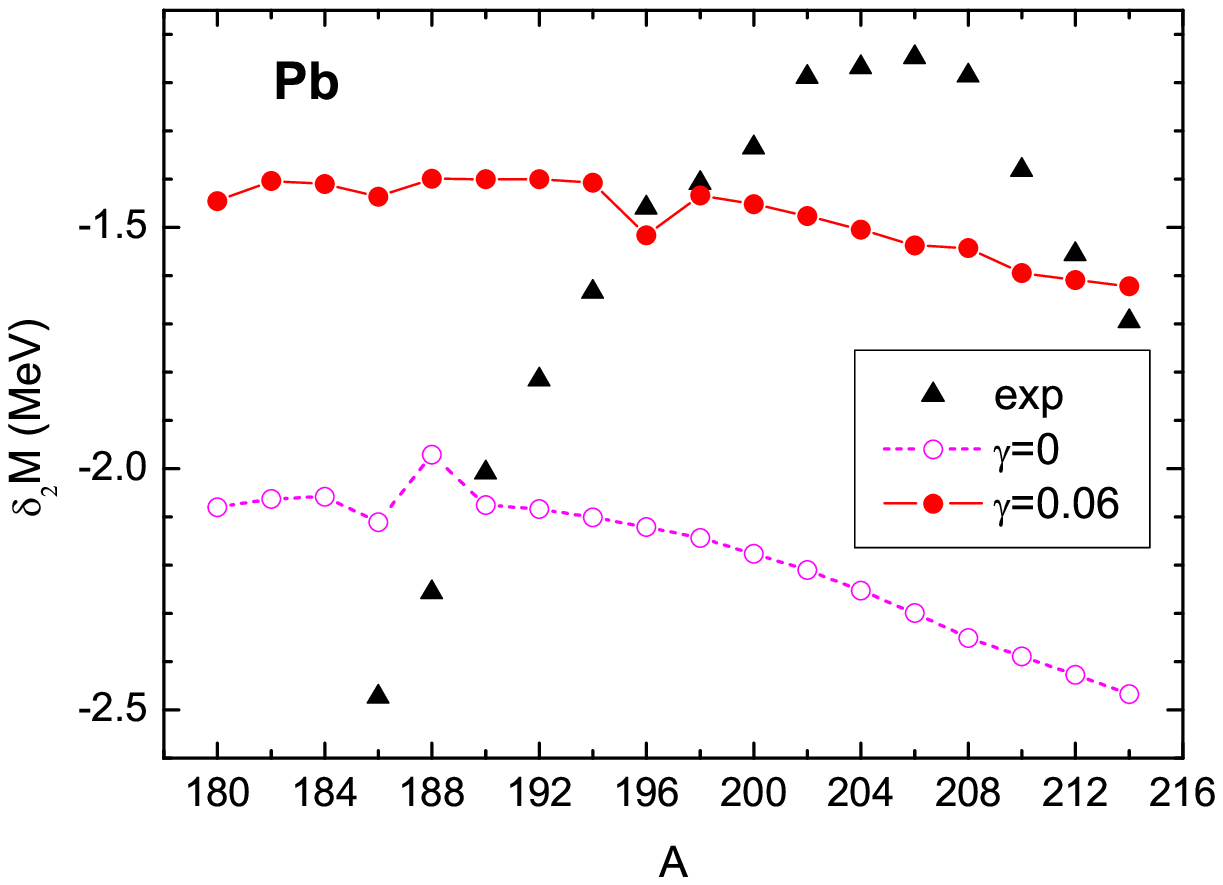}}
\caption{Double mass differences $D_{2p}^+$ for adding to Pb isotopes of two protons}\label{fig:pbD2ppl}
\end{figure}

\begin{figure}
\resizebox{1.0\columnwidth}{!}
{\includegraphics {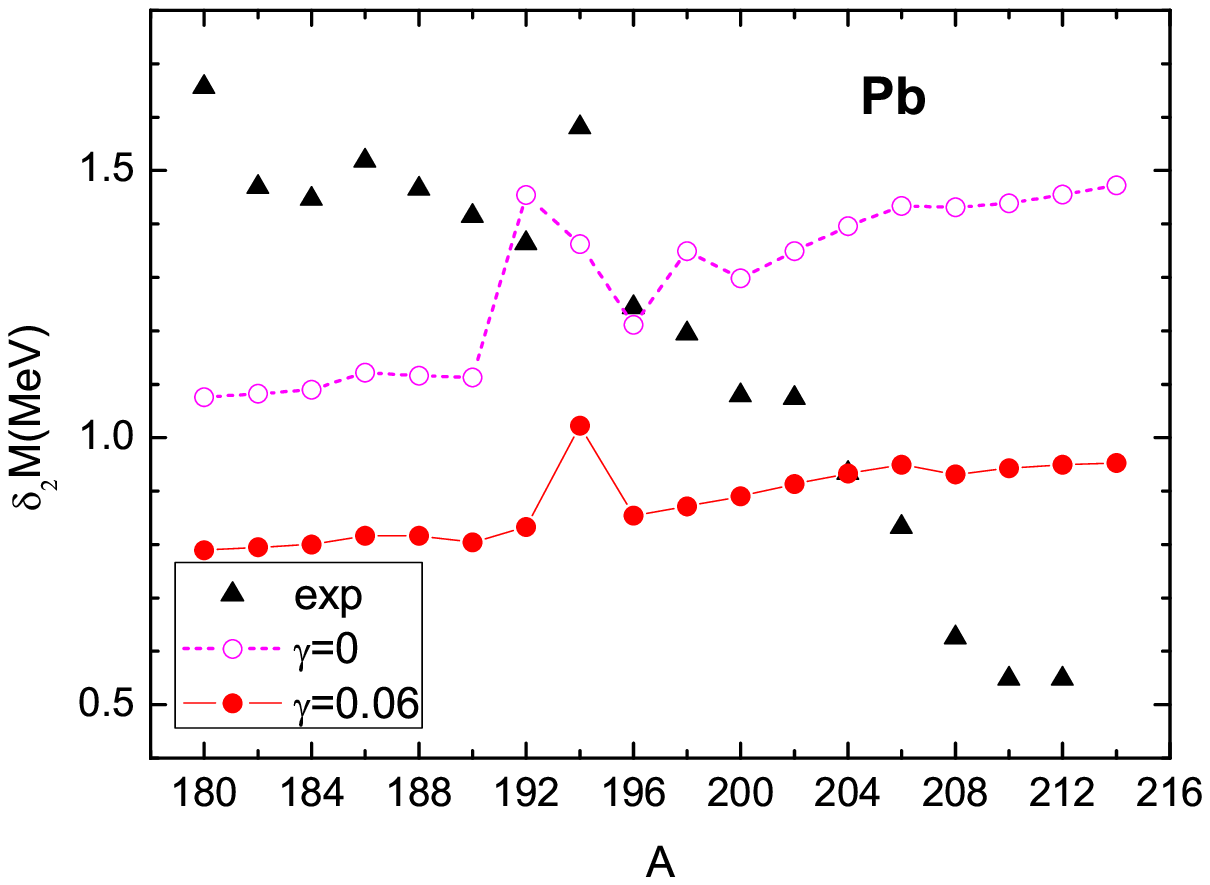}}
\caption{Double mass differences $D_{2p}^-$ for adding to Pb isotopes of two proton holes}\label{fig:pbD2pmi}
\end{figure}

\begin{figure}
\resizebox{0.8\columnwidth}{!}
{\includegraphics {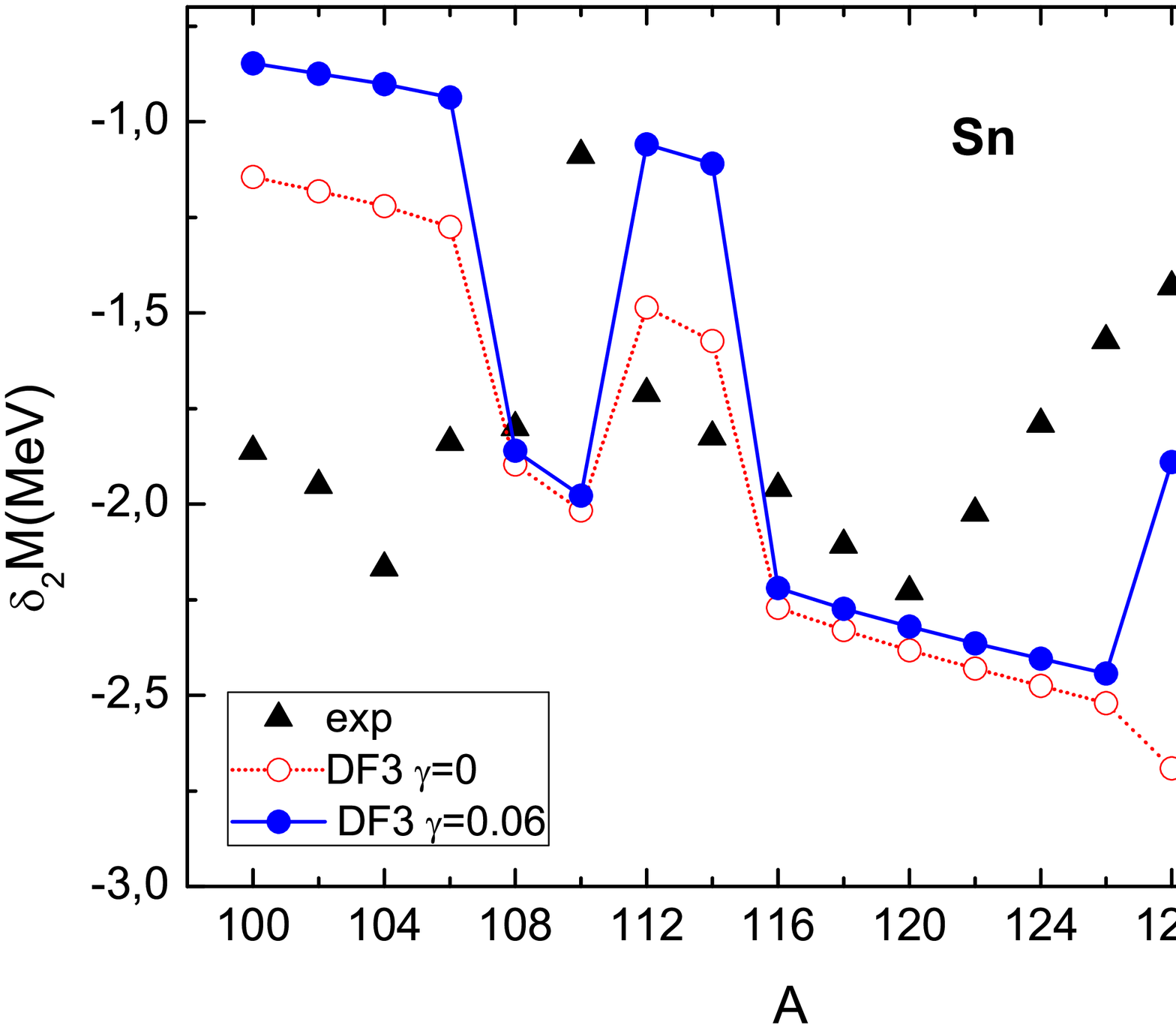}}
\caption{Double mass differences $D_{2p}^+$ for adding to Sn
isotopes of two protons}\label{fig:snD2ppl}
\end{figure}

\begin{figure}
\resizebox{0.8\columnwidth}{!}
{\includegraphics {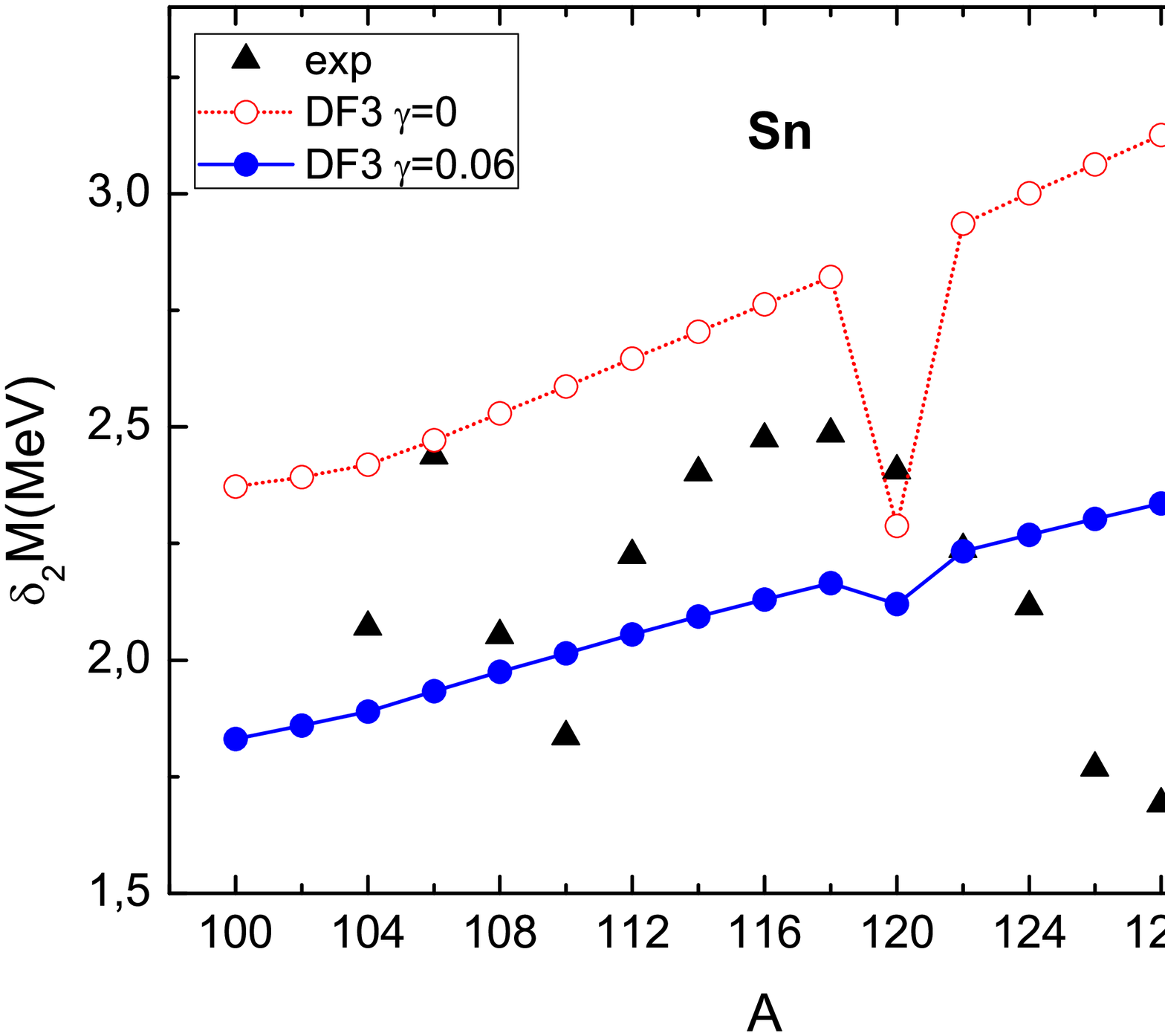}}
\caption{Double mass differences $D_{2p}^-$ for adding to Sn isotopes of two proton holes}\label{fig:snD2pmi}
\end{figure}

\begin{figure}
\resizebox{0.8\columnwidth}{!}
{\includegraphics {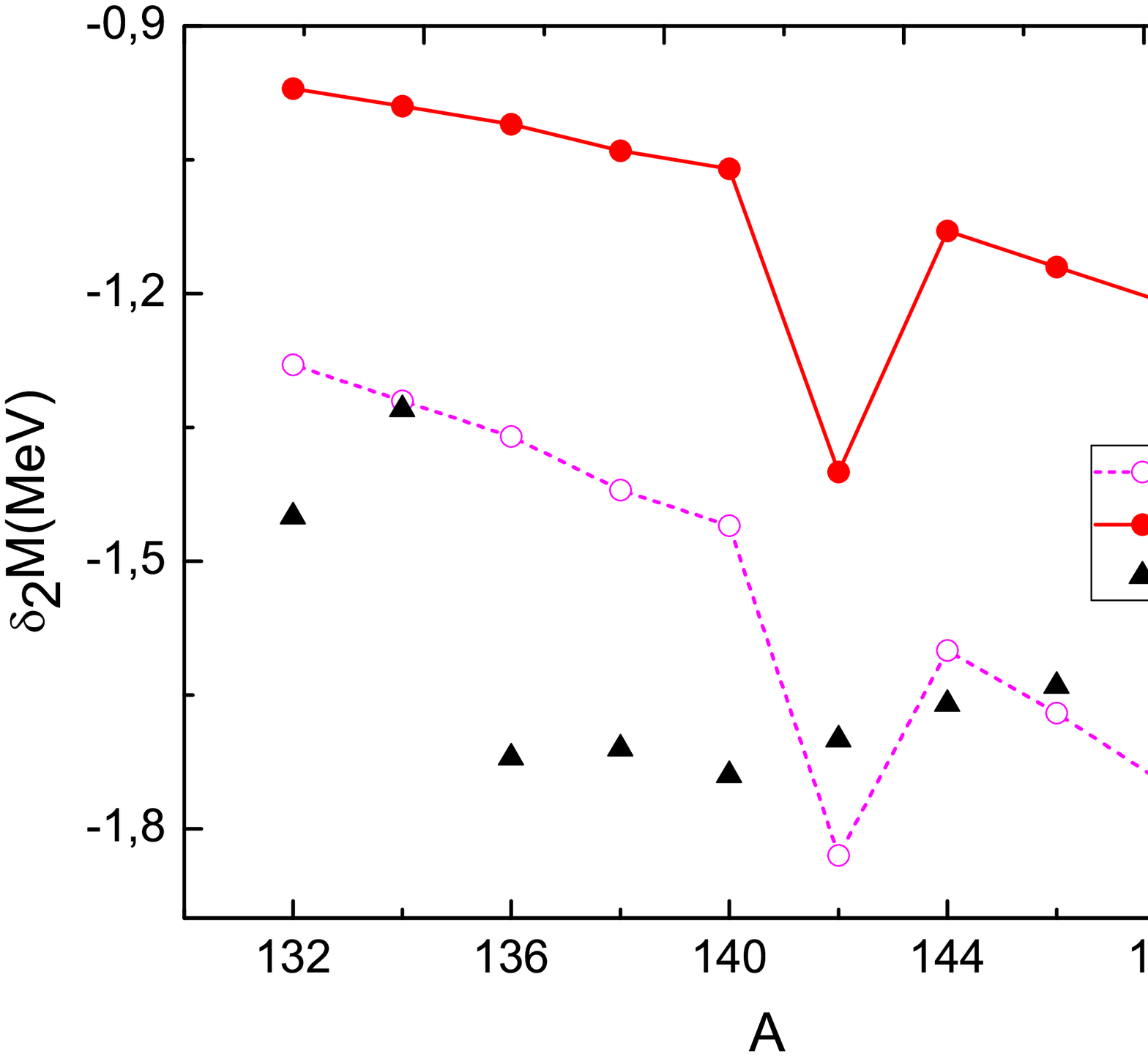}}
\caption{Double mass differences $D_{2n}^+$ for adding to $N=82$ isotones of two neutrons}\label{fig:D2npl}
\end{figure}

\begin{figure}
\resizebox{0.8\columnwidth}{!}
{\includegraphics {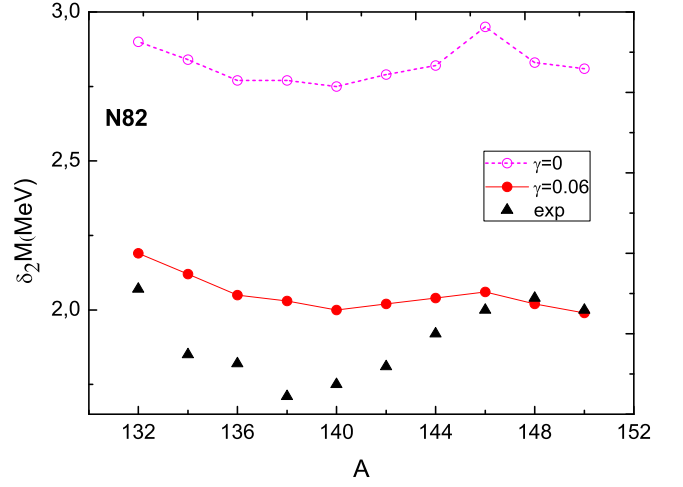}}
\caption{Double mass differences $D_{2n}^-$ for adding to $N=82$ isotones of two neutron holes}\label{fig:D2nmi}
\end{figure}

The set of Eqs. (\ref{Es}),(\ref{Gamma1}) was solved with the functional DF3
only. For  lead isotopes, the proton differences  $D_{2p}^+$  are displayed in Fig. \ref{fig:pbD2ppl}
and the differences  $D_{2p}^-$,   in Fig. \ref{fig:pbD2pmi} and
for the tin isotopes, analogously, in Fig. \ref{fig:snD2ppl} and \ref{fig:snD2pmi}.
In similar way, the neutron double differences
 for isotones $N{=}82$, are displayed in Figs. \ref{fig:D2npl} and \ref{fig:D2nmi}.
Note that the opposite signs of the quantities
$D_2^+$ and $D_2^-$ occur owing to the factor of $(1-2n_1)$ in Eq.
(\ref{Es}): particles attract each other whereas holes, repel.
The  results of the {\it ab initio} calculation ($\gamma{=}0$) and
with the phenomenological addendum ($\gamma{=}0.06$) are shown. To
quantify the accuracy of calculations, just as for the pairing gap in Sect. 2, we calculated the rms
deviations of the theory from experiment,
\beq \label{dD_avr} \overline{\delta M_{2p,2n}} =  \sqrt{\frac 1
{N_{p,n}} \sum_{i} \left(D^{(i)}_{\rm theor}-D^{(i)}_{\rm
exp}\right)^2} \eeq with obvious notation. For the lead isotopes, we find\\ $\overline{\delta
M_{2p}}(\gamma{=}0){=}0.64\;$MeV and $\overline{\delta
M_{2p}}(\gamma{=}0.06){=}0.42\;$MeV, $N_p{=}26$. For the tin isotopes, we get
$\overline{\delta
M_{2p}}(\gamma{=}0){=}0.68\;$MeV and $\overline{\delta
M_{2p}}(\gamma{=}0.06){=}0.45\;$MeV, $N_p{=}21$.
For isotones with
$N{=}82$ we obtain $\overline{\delta M_{2n}}(\gamma{=}0){=}0.66\;$MeV and
$\overline{\delta M_{2n}}(\gamma{=}0.06){=}0.42$ MeV, $N_n{=}19$.
We included into analysis only nuclei with the experimental error of the double
difference less than 0.1 MeV
The total disagreement for neutrons and protons  is $\overline{\delta
M_{\rm tot}}(\gamma{=}0){=}0.66\;$MeV and $\overline{\delta M_{\rm
tot}}(\gamma{=}0.06){=}0.43\;$MeV. Thus, introducing  the phenomenological
addendum makes the agreement with  data better but not so much as for the gap.

One can see that the agreement is worse
for lighter isotopes and isotones which are close to the drip line, the
chemical potential $\mu$ in Eq. (\ref{Es}) being small. The reason is that
in vicinity of drip line the concept of energy independent mean field
becomes erroneous \cite{mu1,mu2}. If we limit the analysis to nuclei
with $\mu{<}{-}4\;$MeV, we obtain $\overline{\delta
M_{2p}}(\gamma{=}0){=}0.53$ MeV and $\overline{\delta
M_{2p}}(\gamma{=}0.06){=}0.27$ MeV, $N_p{=}10$ for lead isotopes and
 $\overline{\delta
M_{2p}}(\gamma{=}0){=}0.64$ MeV and $\overline{\delta
M_{2p}}(\gamma{=}0.06)$ ${=}0.40\;$MeV, $N_p{=}16$ for the tin chain. For isotones
$N{=}82$ we have $\overline{\delta M_{2n}}(\gamma{=}0)=0.71$ MeV and\\
$\overline{\delta M_{2n}}(\gamma{=}0.06)$ = 0.39 MeV, $N_n{=}16$.
It is worth mentioning that for the $D_{2n}^+$ quantity, Fig. \ref{fig:D2npl},
inclusion of the phenomenological addendum makes agreement significantly worse.
The total disagreement for neutrons and protons is now
$\overline{\delta M_{\rm tot}}(\gamma{=}0){=}0.64$ MeV and
$\overline{\delta M_{\rm tot}}(\gamma{=}0.06){=}0.36$ MeV.
Thus, after the inclusion of the phenomenological correction, we obtain
the agreement only a little worse than for the superfluid subsystem.

\begin{figure}
\resizebox{1.0\columnwidth}{!}
{\includegraphics {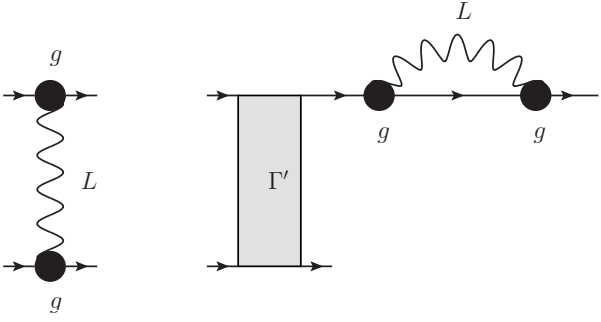}}
\caption{Phonon corrections to the interaction amplitude $\Gamma'$}\label{fig:phon}
\end{figure}

 We observe  that our calculations, which reproduce the value
of $D_2$ on average, do not catch the trend of the $A$-dependence of this
quantity. We can suppose that it could be due to the fact that phonon corrections, included
into the parameter $\gamma$, are not considered explicitly. Estimations show that
for this problem they are not so universal as in the gap problem as far as
the diagonal element  $\left({\cal V}_{\rm
eff}\right)_{11}^{11}$  in Eq. (\ref{Gamma1}) for $\left(
\Gamma'(E_s)\right)_{11}^{11}$ often dominates. In such a situation, the phonon corrections
could strongly depend on the  state $|1\rangle$.

 The main corrections to Eq. (\ref{Es}) caused by the $L$-phonon are shown in diagrams of Fig. \ref{fig:phon}.
 The first one is the so-called induced interaction, the second one is the  ``ends correction''.
 In magic and semi-magic nuclei, the vertex $g$ as a rule can be considered as a small parameter, and
 the so-called  $g^2$-approximation can be used. The $g^2$-corrections are displayed in Fig. \ref{fig:phon}.
 The wavy line corresponds to the phonon ${\cal D}$-function,
${\cal D}(\omega){=}2\omega_L/(\omega^2 -\omega_L^2)$, where
$\omega_L$ is the excitation energy of the  $L$-phonon, and the transferred energy
$\omega{=}0$ for the diagonal matrix element. The vertex
$g(r)$ of creating $L$-phonon and the frequency  $\omega_L$ were found
 \cite{BE2}  by within the self-consistent TFFS
 \cite{KhS}. Note that all  low-lying phonons possess the normal parity
 $\pi{=}(-1)^L$, therefore we do not  show it explicitly. The first diagram can be easily
 evaluated:\beq \delta_{\rm ph}^{(1)}D_2=- \frac
{2(\langle j_1 l_1|| Y_L || j_1 l_1\rangle g_{11})^2}{\omega_L
(2j_1+1)}, \label{del1}\eeq where  $\langle \;|| Y_L || \;\rangle$
denotes the reduced matrix element \cite{BM1}, and $g_{11}$ is the
radial matrix element of $g(r)$.

As to the second diagram, one has the same corrections for
each of the ends. They can be readily summed \cite{AB}. As the result, each of the ends
should be multiplied by
$\sqrt{Z_1}$, where  $Z_1$ is the residue of the Green function $G(\eps)$ at the point
$\eps{=}\eps_1$. The corresponding correction to the double mass difference is equal to  \beq \delta_{\rm ph}^{(2)}D_2= (Z_1^2-1)D_2.
\label{del2}\eeq This expression contains terms of higher order in  $g^2$ but such partial summation
has clear physical meaning. In addition, sometimes the difference
 $Z_1-1$ is not small. The matter is that the ratio
$\bar{g}{=}g_{12}/\delta \eps_{12}$, where $\delta
\eps_{12}{=}\eps_1-\eps_2\pm \omega_L$, is the actual dimensionless ``small'' parameter.
For collective  $L$-phonons the matrix elements
$g_{12}\simeq 1\;$MeV, whereas the typical value of the denominator is
$\delta \eps_{12}\simeq \eps_{\rm F}/A^{1/3}$, $\eps_{\rm F}$ being the Fermi
energy and $A$  the mass number of the nucleus under consideration. Thus the condition
 $\bar{g}\ll 1$ is valid. However, sometimes  the  ``resonance'' cases occur when the quantity
$\delta \eps_{12}$ is anomalously small. Then the $g^2$-approximation does not work and higher order terms
should be summed up. In such cases, it is natural to multiply by $\sqrt{Z_1}$ also the ends of the
first diagram in Fig. \ref{fig:phon}. Then the expression for the double mass difference with phonon correction
is as follows: \beq \tilde{D}_2=\left( D_2 + \delta_{\rm
ph}^{(1)}D_2\right)Z_1^2. \label{rez} \eeq

Let us calculate from this formula the corrections to the proton
double mass difference
$D_{2p}$ first for the double-magic $^{208}$Pb nucleus.
In this nucleus the collective
$3^-$-phonon with the energy $\omega_3{=}2.684\;$MeV plays the main role.
Adding  two proton particles, we have  $D_{2p}^+(\gamma=0){=}-2.35\;$MeV, $|1\rangle{=}1h_{9/2}$, $Z_1{=}0.98$, $\delta_{\rm
ph}^{(1)}D_{2p}{=}{-}0.14\;$MeV and $\tilde{D}_{2p}{=}{-}2.38$ MeV.
Thus, in this case the phonon correction, $\delta_{\rm
ph}D_{2p}{=}$ $\tilde{D}_{2p}{-}D_{2p}{=}{-}0.03\;$MeV is negligible.
Adding  two proton holes, we have  $D_{2p}^-(\gamma=0){=}+1.43\;$MeV, $|1\rangle{=}3s_{1/2}$,
obviously, we have  $\delta_{\rm ph}^{(1)}D_{2p}{=}0$ due to the angular momentum conservation.
Then, we have $Z_1{=}0.96$  and $\tilde{D}_{2p}{=} 1.32\;$MeV, thus the phonon effect
$\delta_{\rm ph}D_{2p}{=}{-}0.11$ MeV is more than for the $1h_{9/2}$ state but also less than  10\%.
Another situation takes place for the semi-magic nucleus  $^{204}$Pb.
Here the phonon corrections appear mainly due to the low-lying
$2^+$-phonon, $\omega_2{=}0.882\;$MeV. In this case, for the state
$1h_{9/2}$ we have $Z_1{=}0.74$, $\delta_{\rm
ph}^{(1)}D_{2p}{=}{-}0.55\;$MeV,  $\tilde{D}_{2p}{=}{-}1.53\;$MeV and
$\delta_{\rm ph}D_{2p}^+{=}0.72\;$MeV. For the state $3s_{1/2}$ again we have
$\delta_{\rm ph}^{(1)}D_{2p}{=}0$, but $Z_1{=}0.82$,
$\tilde{D}_{2p}{=}0.93\;$MeV and $\delta_{\rm
ph}D_{2p}^-{=}{-}0.47$ MeV. Thus, for both states the phonon correction is approximately
of the same magnitude as the effect of the phenomenological addendum in Eq.  (\ref{Vef1}) at $\gamma{=}0.06$.
Hence now we must take smaller  value of  $\gamma$. In particular, for the state
$3s_{1/2}$ the value of $\gamma{=}0$ looks preferable. Evidently, more consistent approach is necessary
with systematic account for the phonon corrections with a new readjustment of   the parameter $\gamma$.

\section{Conclusions}

Two aspects of the problem of evaluating double odd-even mass differences $D_2$ in semi-magic nuclei are
analyzed. A semi-magic nucleus contains two components with different properties, a superfluid subsystem
and a non-superfluid one. For the superfluid subsystem, $D_2$ is supposed to be equal to $2\Delta$,
the gap $\Delta$ being solution of the gap equation. Such equalization contains an inherent  inaccuracy
for $\Delta$ of the order of $\simeq 0.1\div0.2\;$MeV \cite{Pankr3} and an additional one due to
particle number non-conservation effects $\simeq 0.1\;$MeV \cite{part-numb}. For $D_2$ value they should
be multiplied by a factor of 2. In Ref. \cite{Pankr3} the gap equation was solved with the effective
pairing interaction of
the semi-microscopic model   in which the main term found from  first principles is
supplemented with a small phenomenological addendum containing one
phenomenological parameter $\gamma$  supposed to be the same for all medium and heavy atomic
nuclei. The data for 34 nuclei were analyzed for isotopes of the lead and tin chains and isotones
$N=82$. Neighbors of the double magic nuclei were excluded, as for them particle number non-conservation
effects are especially large \cite{part-numb}.
Overall disagreement of the theory with the data is $\sqrt{\overline{(\Delta_{\rm th}-
\Delta_{\rm exp})^2}}{\simeq}$ 0.13 MeV for the DF3 functional \cite{Fay} and
${\simeq}0.14\;$MeV, for the DF3-a one \cite{Tol-Sap}. For the mass differences $D_2$ it corresponds to
accuracy a little better then 0.3 MeV.

For the non-superfluid
subsystem,  $D_2$ is found  solving the equation for two-particle Green function
of normal systems. This equation  contains exactly the same effective
pairing interaction as the gap equation.
Results of calculations for the lead and tin isotopes and $N=82$ isotones
with the DF3 functional and with the same value $\gamma{=}0.06$ of the phenomenological
parameters are presented here in Figs. \ref{fig:pbD2ppl}--\ref{fig:D2nmi}.
It is worth to note that the problem for non-superfluid nuclei,
in principle, does not contain so serious inaccuracies as the pairing problem.
The main approximation which is made is that the two-particle Green function $K$ which
concerns the array of $A, A\pm 1,A\pm 2$ nuclei
is described in the  basis of the $A$-nucleus. Estimation of the accuracy of this approximation
is $\sim 2/A$. However, the agreement turned out to be a little worse than for superfluid case,
$\overline{\delta M_{\rm tot}}{=}0.36\;$MeV. The reason is, evidently, in phonon corrections
which are in this case not so regular as in the gap equation. Estimation of these
corrections shows that, indeed, a more consistent approach is necessary
with systematic account for the phonon corrections, the parameter $\gamma$ being readjusted anew.

\section{Acknowledgment}
We thank S.V. Tolokonnikov for valuable discussions. The work was partly supported by the DFG and RFBR Grants
Nos.436RUS113/994/0-1 and 09-02-91352NNIO-a,
by the Grants NSh-7235.2010.2  and 2.1.1/4540 of the Russian Ministry for Science
and Education, and by the RFBR Grants  11-02-00467-a and 12-02-00955-a.
Two of us (E. S. and S. P.) thank the INFN, Seczione di Catania, for hospitality.

{}

\end{document}